\documentclass[aip,amsmath,amssymb]{revtex4-1}

\usepackage{graphicx}
\usepackage{dcolumn}
\usepackage{bm}

\begin{document}


\title{Electronic Circuit Analog of Synthetic Genetic Networks: Revisited} 



\author{Edward H. Hellen}
\email[]{ehhellen@uncg.edu}
\affiliation{Department of Physics and Astronomy, University of North Carolina Greensboro, 
  Greensboro, NC 27402, USA}

\author{Syamal K. Dana}
\affiliation{CSIR-Indian Institute of Chemical Biology, Kolkata 700032, India}

\date{\today}

\begin{abstract}
Electronic circuits are useful tools for studying potential dynamical behaviors of synthetic genetic networks. The circuit models are complementary to numerical simulations of the networks, especially providing a framework for verification of dynamical behaviors in the presence of intrinsic and extrinsic noise of the electrical systems.  Here we present an improved version of our previous design of an electronic analog of genetic networks that includes the 3-gene Repressilator and we show conversions between model parameters and real circuit component values to mimic the numerical results in experiments. Important features of the circuit design include the incorporation of chemical kinetics representing Hill function inhibition, quorum sensing coupling, and additive noise. Especially, we make a circuit design for a systematic change of initial conditions in experiment, which is critically important for studies of dynamical systems' behavior, particularly, when it shows multistability.  This improved electronic analog of the synthetic genetic network allows us to extend our investigations from an isolated Repressilator to coupled Repressilators and to reveal the dynamical behavior's complexity. 
\end{abstract}


\maketitle 

\section{Introduction}
Synthetic genetic networks provide a potential tool to design useful biological functions targeted to perform specific tasks.\cite{hasty2001,elowitz2010,benenson2012}  The early stages of research in this direction were to envision and understand simple networks which provided the basic components for building more complex functional devices. Emphasis was first given to the design of a genetic toggle switch\cite{gardner2000} and an oscillator known as the Repressilator consisting of a 3-gene inhibitory ring that has been expressed in \textit{E. coli}.\cite{elowitz2000} Later, electronic circuits were suggested and used to study the dynamics of synthetic genetic networks.\cite{mason2004,wagemakers2006,buldu2007,tokuda2010} Electronic circuits, in general, allow precise control of system parameters and provide a minimal set-up for experimenting with a dynamical behavior in the presence of intrinsic and extrinsic noises.  This option is useful, to predict various desired functional behaviors in electronic analogs of synthetic genetic networks which are difficult to control in real biological experiments. 

We have designed electronic circuits, in the past, to model genetic networks configured to investigate dynamical behaviors of the Repressilator\cite{hellen2011, hellen2013} and to perform noise-aided logic operations.\cite{hellen2013b}  In the Repressilator studies, we first considered an isolated Repressilator and verified the functional form of the predicted oscillations.\cite{hellen2011}  Then we incorporated a bacterial-inspired method of quorum sensing (QS) coupling\cite{Garcia-Ojalvo2004} into our Repressilator circuit by adding a feedback chain to the 3-gene inhibitory ring.  This additional pathway led to a rich variety of dynamical behavior, including multistability, for the QS-modified isolated Repressilator.\cite{hellen2013} Simulations of this single Repressilator system have even demonstrated period doubling chaotization.\cite{potapov2012} The next step of allowing the QS mechanism to couple Repressilators together as has been done in simulation\cite{Garcia-Ojalvo2004,ullner2007,ullner2008} proved difficult using our previous circuit models. This difficulty leads us to make improvements of the circuit including a complete redesign of the QS circuitry, which we present here in detail.  The improved design allowed us to investigate the more complex dynamics that exist for coupled Repressilators [in prep] and to access the full QS-parameter range of the mathematical model. Apart from their potential use in synthetic biological devices, coupled Repressilators are of interest because they belong to the field of coupled nonlinear oscillators which is essential for the understanding of a wide variety of biological phenomena.\cite{strogatz1993}  

It is crucial to have a precise control of the initial conditions when studying a multistable system like the QS-coupled Repressilators  so that all of the coexisting attractors for a given set of parameters can be captured.  We describe the use of an analog switch to set the initial conditions by initializing capacitor voltages to the desired values.  Multistability also opens the possibility of noise-induced transitions from one attractor to another. Therefore we use our previous noise circuit\cite{hellen2013b} and the genetic network circuit as a test-bed to demonstrate noise-induced transitions between attractors within the QS-coupled Repressilator system.

We begin with the mathematical model and the analog circuit for the genetic network of Repressilators coupled via QS.  Then we present our circuit analysis to relate the circuit with the mathematical model, use the QS circuit to verify the numerical predictions, and show results for coupled Repressilator circuits.  Finally, we describe how to set the initial conditions and incorporate additive noise in the electronic circuit.
  
\begin{figure}
\includegraphics[width=0.5\textwidth]{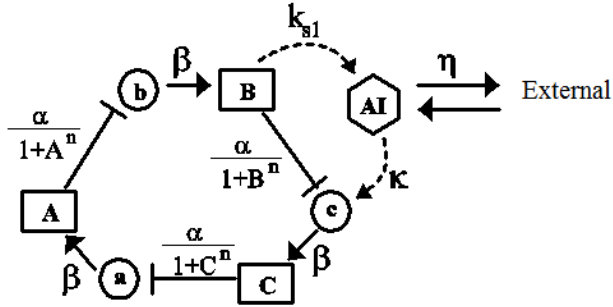}
\caption{\label{fig:rep-qs}Repressilator with quorum sensing feedback. mRNA (a,b,c) and their expressed proteins (A,B,C) form the 3-gene inhibitory loop referred to as the Repressilator. Quorum sensing is provided by the additional feedback loop of the small auto-inducer molecule which can diffuse through the cell membrane thereby exchanging with the external medium. }  
\end{figure}

\section{Model: Repressilator with quorum sensing}
We present here the mathematical model and the circuit model for the genetic network of our interest. The following sections show our analysis which connects the circuits to the equations.  

Figure \ref{fig:rep-qs} shows a Repressilator with a QS feedback loop. The mRNA (a,b,c) and their expressed proteins (A,B,C) form the 3-gene inhibitory loop referred to as the Repressilator.\cite{elowitz2000} It is named Repressilator because each gene's output``represses" the next gene's expression, resulting in stable oscillations of protein concentrations over a very broad interval of parameter values. Thus the 3-gene ring network works as a genetic oscillator. The QS feedback loop uses a small auto-inducer (AI) molecule to provide an indirect activation path from $B$ to $C$ to compete with the direct inhibition.\cite{ullner2008} This network structure generally leads to an anti-phase synchronization of two coupled Repressilators, meaning there is a $180^o$ phase difference between the protein oscillations of the two Repressilators. A different network structure placing the feedback loop from $A$ to $C$ has also been employed,\cite{Garcia-Ojalvo2004} which generally leads to in-phase synchrony.  Interestingly, the network structure does not fully determine the type of synchronization observed between coupled Repressilators as both of these structures are birhythmic--capable of both types of synchrony--depending on the model's parameter values.\cite{potapov2011}  This birhythmic property may be of use in the design of task-oriented devices.  

We use our reduced mathematical model for QS-coupled Repressilators\cite{hellen2013} which is based on previous models\cite{elowitz2000,ullner2008} and applies to the case of fast mRNA kinetics compared to protein kinetics. The model uses standard chemical kinetics $(\beta,\alpha,\kappa,n,k_i)$ including Hill function inhibition, $1/(1+x^n)$, and is 
\begin{subequations}
\begin{align}
\frac{dA}{dt}&=\beta_1\left(-A+\frac{\alpha}{1+C^n}\right)\\
\frac{dB}{dt}&=\beta_2\left(-B+\frac{\alpha}{1+A^n}\right)\\
\frac{dC}{dt}&=\beta_3\left(-C+\frac{\alpha}{1+B^n}+\frac{\kappa S}{1+S}\right)
\label{ode-c}\\
\frac{dS}{dt}&=-k_{s0}S+k_{s1}B-\eta\left(S-S_{ext}\right).
\label{ode-s}
\end{align}
\label{rep-eqns}
\end{subequations}
$(A,B,C)$ are the protein concentrations for the Repressilator, and $S$ is the concentration of the AI molecule. The AI can diffuse (diffusion constant $\eta$) through the cell membrane into the external medium, unlike the proteins which are confined inside the cell.  $S_{ext}$ is the AI concentration in the external medium and is a diluted average of the contributions from all the Repressilators, $S_{ext}=QS_{ave}$, where $Q$ is the dilution factor.  For results presented here we use $k_{s0}=1$, $k_{s1}=0.01$, and $\eta=2$ as taken previously.\cite{ullner2008} 

\begin{figure}
\includegraphics[width=0.5\textwidth]{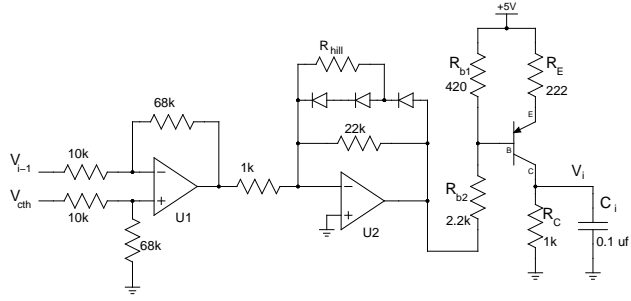}
\caption{\label{fig:gene-circuit}Single-gene circuit. Inhibitory input at
  $V_{i-1}$. Expressed protein concentration is represented by
  $V_{i}$. Dual op-amp is LF412 supplied by $\pm5$ V. The
  \textit{pnp} transistor is 2N3906. Resistor $R_{hill}$ is adjusted to achieve desired Hill-function $n$ value. Capacitor value $C_i=0.1\mu f$ is for $\beta=1$. } 
\end{figure}

The circuit for a single inhibitory gene shown in Fig.~\ref{fig:gene-circuit} is a modification of the previous one.\cite{hellen2011} The transistor current represents the rate of gene expression and the voltage $V_i$ represents the concentration of expressed protein.  $V_{i-1}$ represents the concentration of the repressor, and the $V_{cth}$ adjusts the affinity of the repressor binding to the gene's DNA. The Hill function inhibition in Eq.\ \eqref{rep-eqns} is accounted for by the dependence of the transistor current on repressor concentration voltage $V_{i-1}$. This dependence is derived in the next section. 

\begin{figure}
\includegraphics[width=0.475\textwidth]{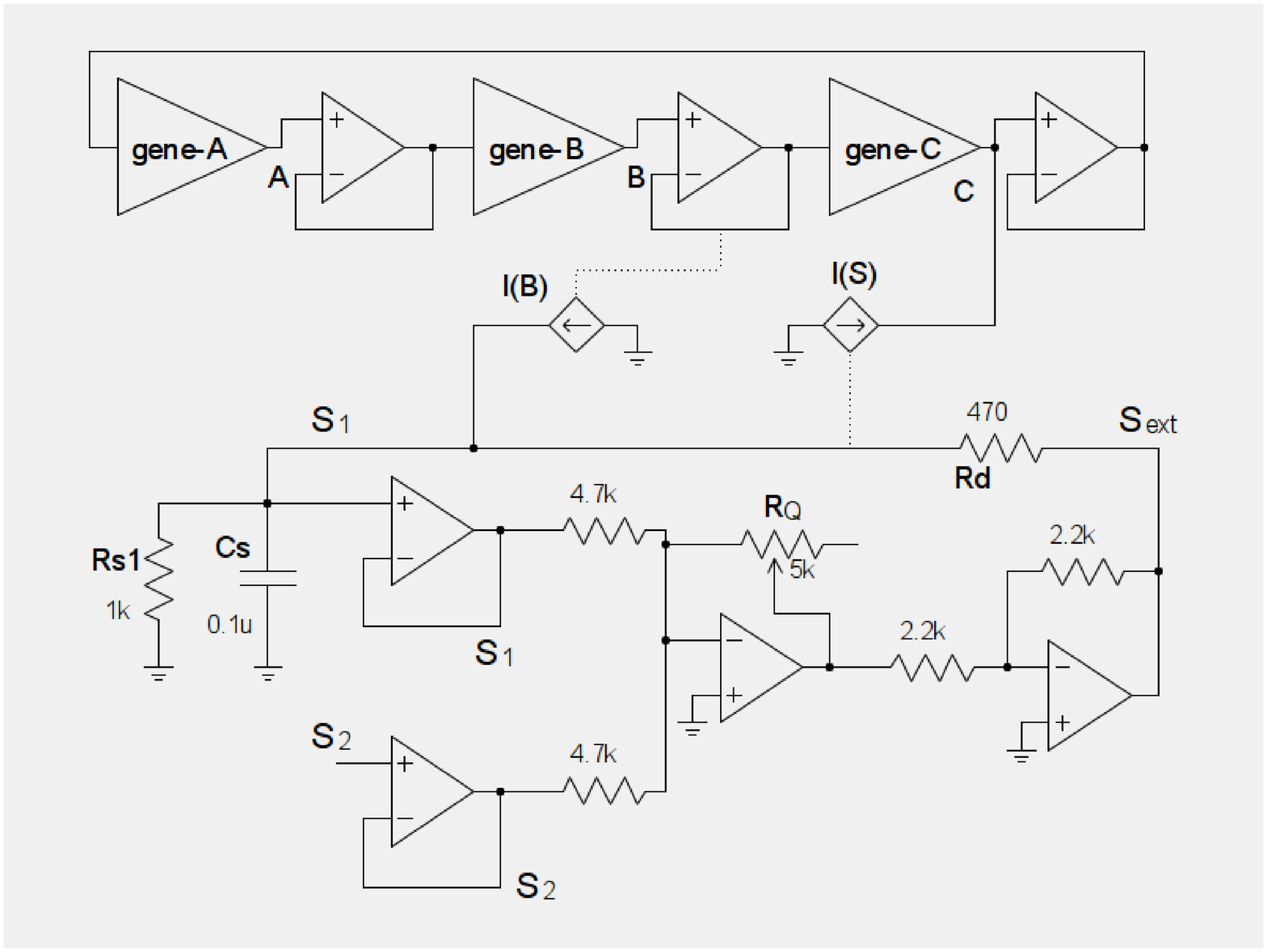}
\caption{\label{fig:rep-circuit}Circuit for Repressilator with QS feedback. The Repressilator consists of the closed ring of genes A, B, and C. The quorum sensing loop is from $B$ through $S_1$ to $C$. Protein $B$ creates auto-inducer $S_1$ via the voltage-controlled current source $I(B)$, and $S_1$ activates production of $C$ via $I(S)$. Each ``gene" triangle corresponds to the single gene circuit in Fig.~\ref{fig:gene-circuit}. $S_2$ is the contribution from a second Repressilator (not shown) and $S_{ext}$ is the auto-inducer concentration in the external medium. The op-amps have low offset voltage (below 0.5 mV).} 
\end{figure}

The circuit for a Repressilator with quorum sensing feedback shown in Fig.~\ref{fig:rep-circuit} is a complete redesign of that presented previously.\cite{hellen2013}  The Repressilator consists of the closed 3-gene loop with op-amp buffers between the genes. The QS circuitry takes input from current source I(B) controlled by the repressilator's $B$-protein voltage, and feeds back to the Repressilator's $C$-protein via source I(S).  The feedback activation in the mathematical model is through the binding-site occupation term $S/(1+S)$.  We show below that the circuit accounts for the activation via QS by using a piece-wise continuous linear behavior, modeled by $\min(0.8S,1)$ and hence we replace Eq.\ \eqref{ode-c} by
\begin{equation}
\frac{dC}{dt}=\beta_3\left(-C+\frac{\alpha}{1+B^n}+\kappa \: \min(0.8S,1)
\right)
\label{new-C-eqn}
\end{equation}
In Fig.~\ref{fig:rep-circuit} $S_1$ is the AI concentration belonging to the shown Repressilator. Coupling this Repressilator to a second Repressilator (not shown) is accomplished by adding their respective AI concentrations, $S_1$ and $S_2$, thus creating $S_{ext}$, the concentration of AI in the external medium.  Figure \ref{fig:rep-circuit} shows the connection of $S_2$ to the op-amp at the bottom of the figure and the combination with $S_1$ to produce $S_{ext}$. 

\subsection{Single Gene Circuit with Hill-function}
We now analyse the circuit for a single gene and show how the inhibitory Hill function behavior is reproduced. In the process we find useful results: how to connect model parameters $n$ and $\alpha$ to circuit parameters, and the minimum accessible value of $n$.  

Applying current conservation to the capacitor voltage in Fig.~\ref{fig:gene-circuit}, and normalizing by a scaling parameter $V_{th}$ gives,
\begin{equation}
R_CC_0\frac{dx_i}{dt}=\frac{C_0}{C_i}\left(-x_i +\frac{I_tR_C}{V_{th}}\right)
\label{gene-circ-eqn}
\end{equation} 
where $x_i=V_i/V_{th}$ is the dimensionless protein concentration and $I_t$ is the transistor's current collector. $R_CC_0$ is the time-scale and it normalizes the time variable, thereby making $t$ dimensionless. A comparison with Eq.\ \eqref{rep-eqns} gives a useful relation between the model parameters and the circuit values,
\begin{equation}
\beta_i=\frac{C_0}{C_i},\: \alpha=\frac{I_{max}R_C}{V_{th}}
\label{beta-alpha} 
\end{equation}
where $I_{max}$ is the maximum transistor current and its relation to $I_t$ is defined below clearly to derive the Hill function behavior in the circuit.  

The gene inhibition in Eq.\ \eqref{rep-eqns} is controlled by the Hill function
\begin{equation}
H(x)=\frac{1}{1+x^n}
\label{Hill}
\end{equation}
where $x$ is the dimensionless inhibitory protein concentration.  The scaling parameter $V_{th}$ accounts for the inhibitor's equilibrium binding constant. Comparing Eqs.\ \eqref{rep-eqns} and \eqref{gene-circ-eqn} shows that the Hill function behavior must be accounted for in the circuit by the transistor current's dependence on input voltage $V_{i-1}$. In this section we derive this current-voltage dependence. The key elements are to get the correct slope at $x=1$ where $H(x=1)=0.5$ and to approximate the Hill function's positive curvature  decay to zero.  

The op-amp U2 in Fig.~\ref{fig:gene-circuit} has different gains, $G_{-2}$ when $V_{i-1} < V_{cth}$, and $G_{+2}$ when $V_{i-1} > V_{cth}$.  For the selected component values in the circuit, the subtraction op-amp U1 has a gain $G_1 = -6.8$, and inverting op-amp U2 has $G_{-2} = -22$ and $G_{+2}$ is an amplitude-dependent diminishing gain due to the three diodes in the feedback for U2.  The diodes create the positive curvature decay of the Hill function.
  
The gene inhibition in the circuit corresponds to $V_{i-1}$ surpassing $V_{cth}$, which causes the output of U2 to go positive and thereby turns off the \textit{pnp} transistor resulting in no current from the collector.  The maximum output voltage of U2 is about 2.0 V when the three diodes are fully conducting in their forward biased state.  The resistors $R_{b1}$ and $R_{b2}$ are chosen such that an output voltage at U2 of 2.0 V causes a drop of $(0.42/2.62)(5 - 2) = 0.48$ V across $R_{b1}$ which is small enough so that the transistor current is essentially zero. Maximal protein expression in the circuit corresponds to $V_{i-1} = 0$ which results in U2 output going negative with a limit at the lower saturation level $V_{-sat} = -3.5$ V for the dual op-amp LF412 supplied with $\pm 5$ V.  We assume that the gain $G_1G_{-2}$ is large enough so that the output of U2 reaches $V_{-sat}$ when $V_{i-1} = 0$.  Later we determine a practical restriction on Hill coefficient $n$ imposed by this assumption.
 
We predict the transistor's collector current in Fig.~\ref{fig:gene-circuit} when the output of U2 varies between -3.5 and 2.0 V.  The collector current is essentially the current in $R_E$ since the transistor is in the active region. The voltage across $R_{b1}$ is $f(5-G\Delta V)$ where the fraction $f = 0.42/2.62 = 0.160$ is the voltage divider gain, $\Delta V = (V_{i-1}-V_{cth})$, and $G$ is the overall gain of the 2 op-amps.  The current in $R_E$, and therefore the transistor current, is  
\begin{equation}
I_t=\frac{f(5-G\Delta V)-V_{eb}}{R_E}
\end{equation}
where $V_{eb}$ is the emitter-base voltage.  $V_{eb}$ varies from about 0.5 V when there is essentially zero transistor current ($G\Delta V\approx 2$ V) to a maximum of about $V_{ebmx}=0.70$ V at maximum current ($G\Delta V=V_{-sat}$).  Maximal protein expression occurs for $V_{i-1} = 0$ (no inhibition) and thus $G\Delta V = V_{-sat}$ giving the maximum transistor current 
\begin{equation}
I_{max}=\frac{f(5-V_{-sat})-V_{ebmx}}{R_E}.
\label{imax}
\end{equation}
For our chosen circuit components we measure $I_{max}=2.95$ mA and $V_{ebmx}=0.70$ V. This agrees well with the prediction using the large-signal transistor model with saturation current $I_S=7$ fA (which we measured for the 2N3906 transistors), $V_{ebmx}=V_T\ln(I/I_S)=0.026\ln(2.95\textrm{ mA}/7\textrm{ fA})=0.696$ V.  The resulting voltage drop across $R_C$ is easily measured by setting $V_{i-1}=0$, and agrees with that predicted by Eq.\ \eqref{imax} flowing into $R_C=1$ k$\Omega$, $I_{max}R_C= 2.97$ V.   

In the circuit, the Hill function Eq.\ \eqref{Hill} corresponds to the normalized transistor current 
\begin{equation}
\frac{I_t}{I_{max}}=\frac{f(5-G\Delta V)-V_{eb}}{f(5-V_{-sat})-V_{ebmx}}.
\label{it-norm}
\end{equation}
As presented previously,\cite{hellen2011} the circuit approximation of the Hill function is accomplished by setting the slope of the normalized current equal to the slope of the Hill function $dH/dx$ at $x = 1$.  Setting the slopes of Eqs.\ \eqref{Hill} and \eqref{it-norm} equal, using $\Delta V = V_{th}(x_{i-1} - V_{cth}/V_{th})$ with $x_{i-1} = 1$, gain $G = G_1G_{-2}$, Eq.\ \eqref{imax}, 
and $V_{th} = I_{max}R_C/\alpha$ provides a useful result connecting important model parameters $n$ and $\alpha$ to circuit parameters.
\begin{equation}
n\alpha=\frac{4fR_CG_1G_{-2}}{R_E}
\label{n-alpha}
\end{equation}
Using our circuit values $f=0.160$, $R_C=1$ $k\Omega$, and $R_E = 222\Omega$, we determine
$n\alpha=2.88G_1G_{-2}$. Equation \eqref{n-alpha} allows desired model parameters $n$ and $\alpha$ to be achieved in the circuit by adjusting gains $G_1$ and $G_{-2}$. 

\begin{figure}
\includegraphics[width=0.475\textwidth]{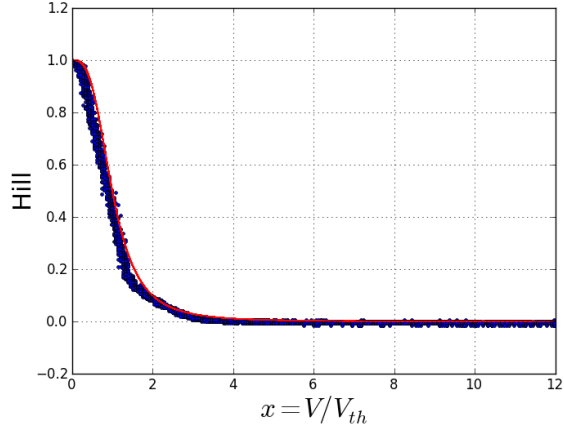}
\caption{\label{fig:Hill}Hill inhibition approximation for the single gene circuit in Fig.~\ref{fig:gene-circuit}.  Numerical Hill inhibition (solid red) and experimentally measured (blue) normalized transistor current. $n=3.2, \alpha=218, R_{hill}=4 k\Omega$. Data was collected with capacitor $C_i$ removed.} 
\end{figure}
 
Next we find the relationship between the binding constant scaling voltage $V_{th}$ and the circuit value $V_{cth}$.  At $x = 1$ the Hill function has a value of 0.5.  The corresponding condition for the circuit is that the normalized transistor current be 0.5 when $V_{i-1} = V_{th}$.  By setting Eq.\ \eqref{it-norm} equal to 0.5, letting $\Delta V = (V_{th} - V_{cth})$ and solving gives   
\begin{equation}
V_{cth}=V_{th}+\frac{\left(2V_{eb}-V_{ebmx}-f\left(5+V_{-sat}\right)\right)}{2fG_1G_{-2}}.
\label{Vcth}
\end{equation}
$V_{eb}$ at half the maximal current is predicted by using 1.5 mA for the transistor current resulting in $V_{eb}=V_T\ln(I/I_S)=0.026\ln(1.5\textrm{ mA}/7\textrm{ fA})=0.678$ V. For the circuit in Fig.~\ref{fig:gene-circuit}, $G_1G_{-2} = (-6.8)×(-22)$, $f=0.160$, $V_{-sat}=-3.5$ V, and using $V_{eb}=0.68$ V and $V_{ebmx}=0.70$ V gives $V_{cth} = V_{th} + 8.8$ mV.  

Figure \ref{fig:Hill} shows the measured approximation of the Hill inhibition for the single gene circuit of Fig.~\ref{fig:gene-circuit} for $n=3.2$, $\alpha=218$, and $R_{hill}=4k\Omega$. The blue dots are the normalized output voltage $V_i/V_{th}$ as a function of normalized input voltage $V_{i-1}/V_{th}$.  It is apparent that as the input voltage surpasses $V_{th}$ (at $x=1$) the transistor current shuts off, closely following the numerically plotted Hill function (solid red line). The location (at $x=1$) and slope of the drop are set by Eqs.\ \eqref{n-alpha} and \eqref{Vcth}, but the positive curvature decay to zero is controlled by $R_{hill}$ in Fig.~\ref{fig:gene-circuit}.  The value of $R_{hill}$ is varied to match the transistor current's decay to that of the Hill function. Our previous circuit model for a single gene\cite{hellen2011} used a piecewise-linear approximation to the Hill function and therefore did not include a positive curvature decay to zero.  

The assumption that the output of op-amp U2 is saturated at $V_{-sat}$ when $V_{i-1} = 0$ (no inhibition) means that $G_1G_{-2}V_{cth} > -V_{-sat}$.  Using the relations between $V_{cth}$ and $V_{th}$ (Eq.\ \eqref{Vcth}), between $V_{th}$ and $\alpha$ (Eq.\ \eqref{beta-alpha}), and between $G_1G_{-2}$ and $n\alpha$ (Eq.\ \eqref{n-alpha}), we find the restriction on the Hill coefficient 
\begin{equation}
n>\frac{2\left(f(5-V_{-sat})-2V_{be}+V_{bemx}\right)}{f(5-V_{-sat})-V_{bemx}}.
\end{equation}
For our circuit values this gives a minimum Hill coefficient of $n=2.12$. This restriction is generally not a problem since the Repressilator in Eq.\ \eqref{rep-eqns} for $\kappa=0$ has a stable fixed point and therefore is not an oscillator for $n<2$ over a wide range of $\alpha$ and identical $\beta$.    

\begin{figure} 
\includegraphics[width=0.5\textwidth]{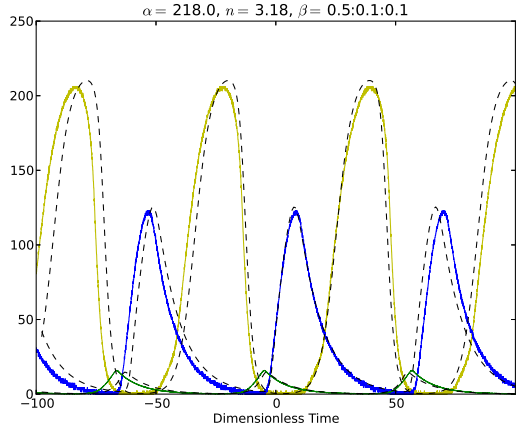}
\caption{\label{fig:rep-time}Repressilator time series. Numerical (dashed) and circuit measurements (colored) for Repressillator with no quorum sensing ($\kappa = 0$ in Eq.\ \eqref{rep-eqns}). $n=3.2, \alpha=218, R_{hill}=4 k\Omega$. } 
\end{figure}

The Repressilator consisting of the 3-gene ring in Fig.~\ref{fig:rep-qs} is modeled by connecting three single-gene circuits in a closed loop depicted by the 3 gene-triangles (A,B,C) in Fig.~\ref{fig:rep-circuit}. Figure \ref{fig:rep-time} shows the measured time series and simulations (dashed lines) for a Repressilator demonstrating the stable protein oscillations ($A,B,C$) with different amplitudes that occur for different protein time-scales $\beta_1=0.5$, $\beta_2=0.1$, and $\beta_3=0.1$.  

\begin{figure}
\includegraphics[width=0.3\textwidth]{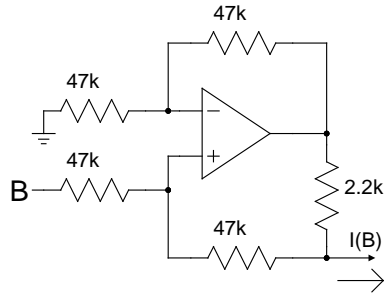}
\caption{\label{fig:B-current}Voltage controlled current source I(B). Protein $B$ voltage controls the current source to the $S$-voltage. $I(B)=V_B/2.2k\Omega$. Op-amp has low offset voltage (below 0.5 mV).} 
\end{figure}
  
\subsection{Circuit for Repressilator with Quorum Sensing}
\begin{figure}
\includegraphics[width=0.475\textwidth]{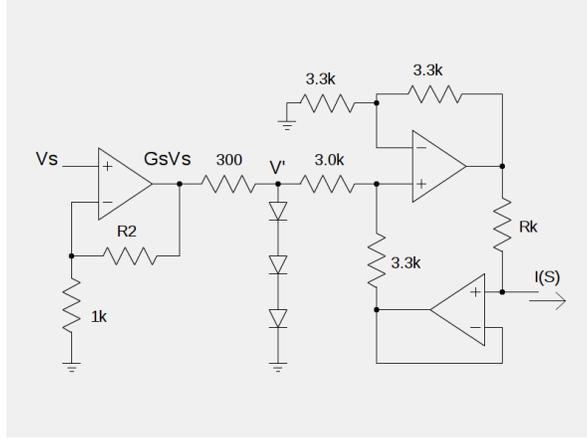}
\caption{\label{fig:S-current}Voltage controlled current source I(S). Auto-inducer $S$-voltage controls the current source feeding back to protein $C$-voltage. For small $S$ the diodes are not conducting and $I(S)=G_SV_S/R_k$. For large $S$ the diodes are forward biased and the voltage $V'$ remains close to 2 V causing the output $I(S)$ to level off. Op-amps have low offset voltage (below 0.5 mV).}  
\end{figure}

Figure \ref{fig:rep-circuit} shows the circuit for a Repressilator with QS feedback.  The circuit is a modification of the earlier version.\cite{hellen2013}  The feedback from $B$ through the current source $I(B)$ to $S_1$, then through $I(S)$ to $C$ corresponds to the AI feedback loop between $B$ and $c$ in Fig.~\ref{fig:rep-qs}. We analyse the circuit to derive relations between the mathematical model and circuit values. Figures \ref{fig:B-current} and \ref{fig:S-current} show the circuits for the voltage dependent current sources $I(B)$ and $I(S)$ used in Fig.~\ref{fig:rep-circuit}.    

The circuit equation corresponding to Eq.\ \eqref{ode-s} comes from circuit analysis for the voltage $V_S$ across the capacitor $C_S$ in Fig.~\ref{fig:rep-circuit}
\begin{equation}
R_{S1}C_S\frac{dV_{S1}}{dt}=-V_{S1}+R_{S1}I(B)-\frac{R_{S1}}{R_d}\left(V_{S1}-V_{ext}\right).
\end{equation}  
$V_{S1}$ and $V_{ext}$ correspond to the scaled voltages $S_1$ and $S_{ext}$ in Fig.~\ref{fig:rep-circuit}. Multiplying both sides by $k_{S0}$, setting $k_{S0}R_{S1}C_S$ to be the same as the time-scale $R_CC_0$ defined for the single-gene circuit, using $I(B)=V_B/2.2k\Omega$ from Fig.~\ref{fig:B-current}, $V_B=V_{th}B$, and dividing by a scaling factor $V_{sth}$ gives
\begin{equation}
\frac{dS_1}{dt}=-k_{S0}S_1+k_{S0}\frac{R_{S1}}{2.2k}\frac{V_{th}}{V_{sth}}B-k_{S0}\frac{R_{S1}}{R_d}\left(S_1-S_{ext}\right)
\end{equation} 
where $V_S=V_{sth}S$ and $V_{ext}=V_{sth}S_{ext}$. Comparison with Eq.\ \eqref{ode-s} gives relations for the activation rate $k_{S1}$ of auto-inducer and the membrane diffusion parameter $\eta$.
\begin{equation}
k_{S1}=k_{S0}\frac{R_{S1}V_{th}}{2.2kV_{sth}},\:\eta=k_{S0}\frac{R_{S1}}{R_d}
\label{Vsth}
\end{equation}
Equation \eqref{Vsth} sets the scaling factor $V_{sth}$.

The equation for the protein $C$ voltage is found in the same way as Eq.\ \eqref{gene-circ-eqn} with the addition of the current $I(S)$ from the feedback loop in Fig.~\ref{fig:rep-circuit}. 
\begin{equation}
R_CC_0\frac{dC}{dt}=\frac{C_0}{C_3}\left(-C+\frac{I_tR_C}{V_{th}}+\frac{I(S)R_C}{V_{th}}\right)
\end{equation}
Comparison with Eq.\ \eqref{new-C-eqn} shows that 
\begin{equation}
\frac{I(S)R_C}{V_{th}}=\kappa \:\min(0.8S,1)
\label{kappa-act}
\end{equation} 
Equation \eqref{kappa-act} imposes two constraints.  First, the maximum value of $I(S)$ must correspond to the right-hand-side maximum $\kappa$ occurring for $S\geq 1.25$, giving 
\begin{equation}
I(S\geq 1.25)\equiv I_{Smax}=\frac{\kappa V_{th}}{R_C}.
\label{ISmax-kap}
\end{equation} 
The maximum current is implemented by adjusting the gain $G_S$ in Fig.~\ref{fig:S-current} so that the $S=1.25$ input voltage $V_S=1.25V_{sth}$ creates a current of 1 mA in the series diodes causing $V'=3\times 0.63=1.9$ V.  The required op-amp output is $G_S(1.25)V_{sth}\approx 1.9+(1mA)(300\Omega)=2.2$ V which provides the appropriate value for $G_S$. Secondly, for currents below the maximum value, Eq.\ \eqref{kappa-act}'s slopes must be the same. From Fig.~\ref{fig:S-current} the current source is $I(S)=G_SV_S/R_\kappa$. For currents below $I_{Smax}$ we use Eq.\ \eqref{kappa-act} and the relation for $V_{sth}$ in Eq.\ \eqref{Vsth} to find the relation between model parameter $\kappa$ and circuit value $R_\kappa$,
\begin{equation}
R\kappa =k_{s0}\frac{G_SR_CR_S}{0.8(2.2k\Omega )k_{s1}\kappa}.
\label{kapa}
\end{equation}
All the values on the right-hand-side except $\kappa$ have been previously determined, therefore Eq.\ \eqref{kapa} provides a direct link between parameter $\kappa$ and circuit value $R_\kappa$. For the values used here the result is $R_\kappa=(56.8G_S)/\kappa$ in $k\Omega$.  

\begin{figure}
\includegraphics[width=0.475\textwidth]{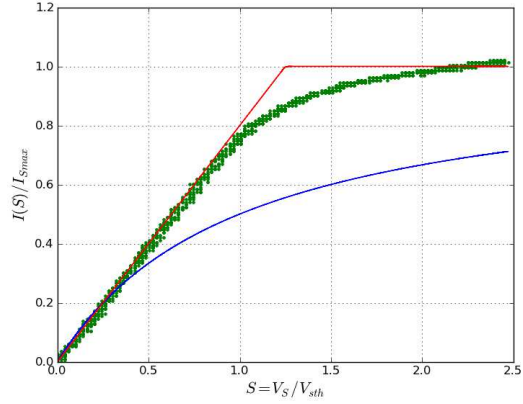}
\caption{\label{fig:S-function}Measured normalized AI activated current(green) and models; piece-wise-linear $\text{min}(0.8S,1)$ (red) and hyperbola $S/(1+S)$ (blue). For $\kappa=21, \alpha=135$.}
\end{figure}
Figure \ref{fig:S-function} shows the measured normalized current $I(S)/I_{Smax}$ from the circuit in Fig.~\ref{fig:S-current}, the piece-wise-linear model $\min(0.8S,1)$ (used in Eq.\ \ref{new-C-eqn}), and the hyperbola $S/(1+S)$ (Eq.\ \ref{ode-c}) for $\kappa=21.3$, $V_{th}=0.0219$ V, and gain $G_s=1.76$. The piece-wise-linear function intersects the hyperbola at $S=0$ and $0.25$. 

We now consider the circuit which creates the AI concentration in the external medium $S_{ext}$.  Each Repressilator circuit contributes its intracellular AI concentration $S_i$ to the external concentration $S_{ext}$. Figure \ref{fig:rep-circuit} shows how two Repressilators are coupled by concentrations $S_1$ and $S_2$ combining to produce 
\begin{equation}
S_{ext}=\frac{2R_Q}{4.7k}S_{ave}=QS_{ave}
\end{equation}
where $S_{ave}= (S_1+S_2)/2$. $Q=2R_Q/4.7k\Omega$ is a dilution factor which in a biological setting ranges from 0 to 1.  For purposes of exploring dynamics in the full parameter range of the mathematical system, we use a $5k\Omega$ potentiometer for $R_Q$ so that we can vary $Q$ from 0 to 2. Our previous circuit design\cite{hellen2013} limited $Q$ variation from 0 to 1.  

\subsection{Selection of Circuit Values}
Here we summarize the practical results for choosing circuit values in Figs.\ \ref{fig:gene-circuit} and \ref{fig:rep-circuit}.  The model parameters are $n$, $\alpha$, $\beta 's$, $\kappa$, $k_{s1}$, and $\eta$.  Some circuit values are chosen independent of the model parameters.  We choose $R_C=R_S=1k\Omega$ and $C_0=C_S=0.1 \mu f$ for characteristic time $0.10$ ms, $R_E=222\Omega$, $V_{-sat}=-3.5$ V (for the LF412 op-amp powered by $\pm 5$ V), and the voltage divider fraction ($R_{b1}$ and $R_{b2}$) in Fig.~\ref{fig:gene-circuit} as $f=420/2620=0.160$.  Resulting measured quantities for the transistor are $I_{max}=2.95$ mA at $V_{ebmx}=0.70$ V, and 1.5 mA at $V_{eb}=0.68$ V.  These currents were shown to be consistent with predictions using the standard transistor model $I(V_{eb})=I_S\exp (V_{eb}/V_T)$.

For Fig.~\ref{fig:gene-circuit}, Eq.\ \eqref{beta-alpha} gives $V_{th}$ and $C_i$, Eq.\ \eqref{n-alpha} gives overall gain $G_1G_{-2}$, and Eq.\ \eqref{Vcth} gives $V_{cth}$. For Fig.~\ref{fig:rep-circuit}, Eq.\ \eqref{kapa} gives $R_\kappa$, op-amp gain $G_S=0.8\times2.2V/V_{sth}$, where $V_{sth}$ is given by Eq.\ \eqref{Vsth}.  The only circuit value not determined by the model parameters is $R_{hill}$ in Fig.~\ref{fig:gene-circuit}. It is convenient to incorporate trim-pots into $R_{hill}$ to adjust the Hill function's positive curvature decay to zero.   

For many choices of parameters the AI concentration $S$ stays below 1, in which case the $S$ activation term $\min(0.8S,1) \rightarrow 0.8S$ meaning there is no need to amplify $V_S$ to impose saturation of $I(S)$. Thus, the current source $I(S)$ in Fig.~\ref{fig:S-current} can be simplified by leaving out the non-inverting op-amp at the input and the 3 diodes, so that $V_S$ connects directly to the $300+3k=3.3k\Omega$.  In this case $G_S=1$ in Eq.\ \eqref{kapa}.    

\subsection{Setting Initial Conditions}
The ability to set initial conditions is crucial when studying systems with multistability so that all attractors can be captured.  We use the 4066 quad analog switch to impose initial conditions by momentarily connecting ``protein" capacitor voltages to desired initial values set by trim-pot voltage dividers with op-amp followers. The 4066 is gated by the output of a 555 timer controlled by a push-button momentary switch (circuit not shown). Improved performance of the 4066 switch is achieved by powering it with 0 and +15 V, compared to the synthetic genetic network circuits powered by $\pm 5$ V.  

\subsection{Other Design Considerations} 
The inexpensive 2N3906 \textit{pnp} transistors used in the gene circuits were selected from a large batch to have nearly the same saturation current, $I_S=7\pm1$ fA, by performing in-house measurements. 

For the case of coupled Repressilator circuits, care was taken to distribute the $\pm 5$ V power rails and ground paths symmetrically to both Repressilators.  The measured voltage difference during operation between respective rails and respective grounds of the two Repressilators was less than 1 mV.   

\section{Measurments: Quorum Sensing Circuit}
We now present experimental results incorporating the new QS circuitry.  We begin with a single Repressilator with QS feedback, followed by two coupled Repressilators. 

\begin{figure}
\includegraphics[width=0.4\textwidth]{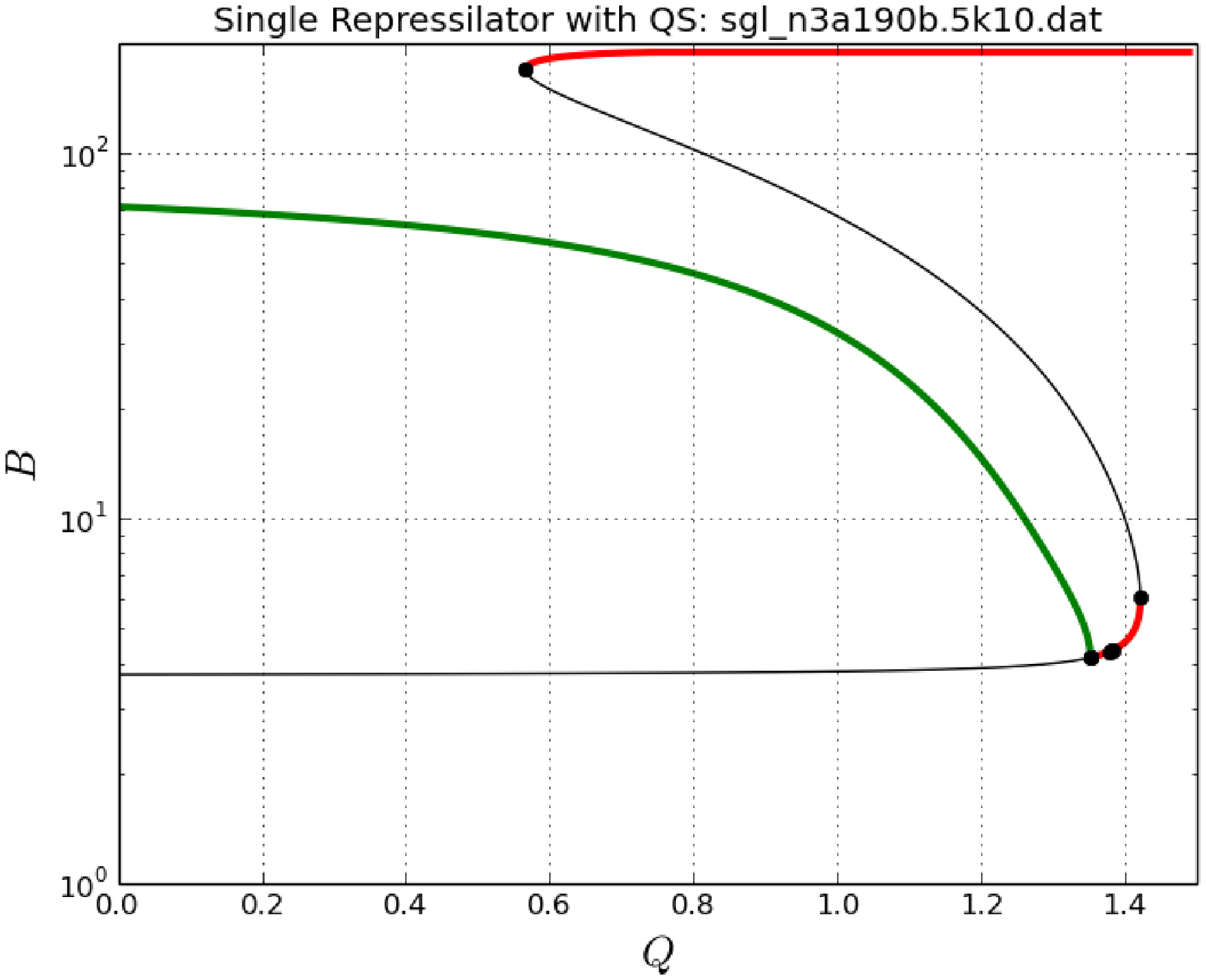}
\includegraphics[width=0.4\textwidth]{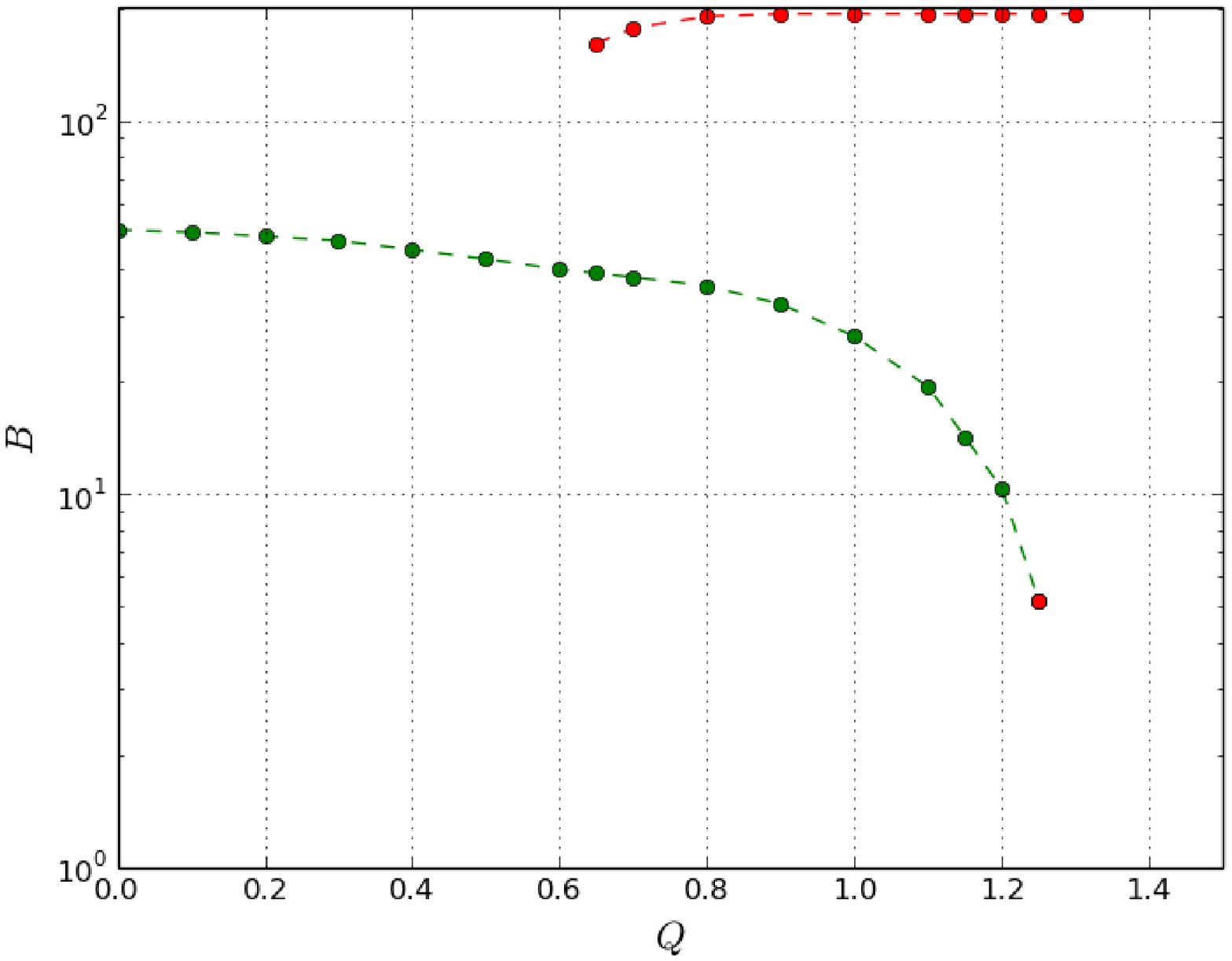}
\caption{\label{fig:Q-bif-sim}Numerical (top) and measured (bottom) $Q$-continuation bifurcation diagrams showing amplitude of protein $B$ for a single Repressilator with quorum sensing. Stable (red) and unstable (black) steady-state.  Limit cycle (green). }
\end{figure}

The case of a single Repressilator with QS feedback corresponds to setting $S_{ext}=QS$ in Eq.\ \eqref{ode-s}.  Measured results from the QS circuit are compared to predictions from numerical simulations using the XPPAUT software.\cite{ermentrout} The desired goal is that the circuit and the simulations have the same structure of dynamical behaviors. A convenient way to do this dynamical comparison is to compare their $Q$-continuation bifurcation diagrams shown in Fig.~\ref{fig:Q-bif-sim}.  These diagrams show the possible amplitudes of $B$ for different $Q$-values.  Steady-state (SS) is either stable (red) or unstable (black), and the limit cycle (LC) oscillations are stable (green).  The $B$-values for the circuit were obtained by normalizing the measured voltage amplitudes by $V_{th}=15.5$ mV which corresponds to the parameter values used in the simulation; $n=3.0$, $\alpha=190$, $\beta_i=0.5,0.1,0.1$, and $\kappa=10$. $R_{hill}=2.7k\Omega$.

\begin{figure}
\includegraphics[width=0.475\textwidth]{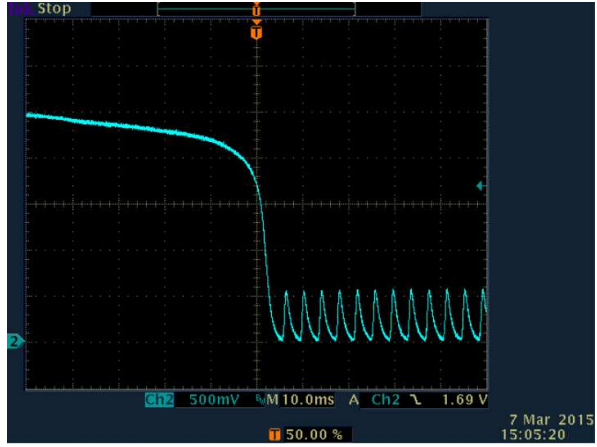}
\caption{\label{fig:SS_to_LC}Screenshot (oscilloscope) of high-$B$-SS to LC transition caused by decreasing $Q$ at 0.64 for a single Repressilator circuit with QS feedback. Protein $B$ voltage shown. Parameters: $n=3$, $\alpha=190$, $\kappa=10$, $R_{hill}=2.7$ k$\Omega$.} 
\end{figure}

In both simulation and circuit measurements Fig.~\ref{fig:Q-bif-sim} shows that increasing $Q$ causes the LC to decrease in amplitude until reaching the low-$B$-SS, and there is coexistence of high-$B$-SS and LC over a broad range of $Q$-values from approximately 0.6 to 1.3.  Both bifurcation diagrams predict that decreasing $Q$ will cause a transition to LC for a system starting from the high-$B$-SS. Figure \ref{fig:SS_to_LC} shows an oscilloscope screenshot of this $Q$-induced high-$B$-SS to LC transition when $Q$ was slowly decreased by adjusting the trim-pot in Fig.~\ref{fig:rep-circuit}.  The transition occurred at a value of $1.5k\Omega$ corresponding to $Q=2\times 1.5/4.7=0.64$ agreeing well with the left-side endpoint of the high-$B$-SS in the bifurcation diagrams. 

The agreement between the circuit and simulation results is not exact in Fig.~\ref{fig:Q-bif-sim}, however, the qualitative structure and relative location of dynamical behaviors are the same.  For the circuit the low-$B$-SS was stable over a $Q$-range narrower than the resolution of $Q$-values and therefore appears as a single data point at the end of the LC-branch. We note that the simulations are able to find the unstable SS (black lines in Fig.~\ref{fig:Q-bif-sim}), whereas the circuit, of course, can only find stable dynamics. We conclude that the quorum sensing circuit achieves the goal of having the same dynamical behavior as the mathematical model.  

\begin{figure}
\includegraphics[width=0.4\textwidth]{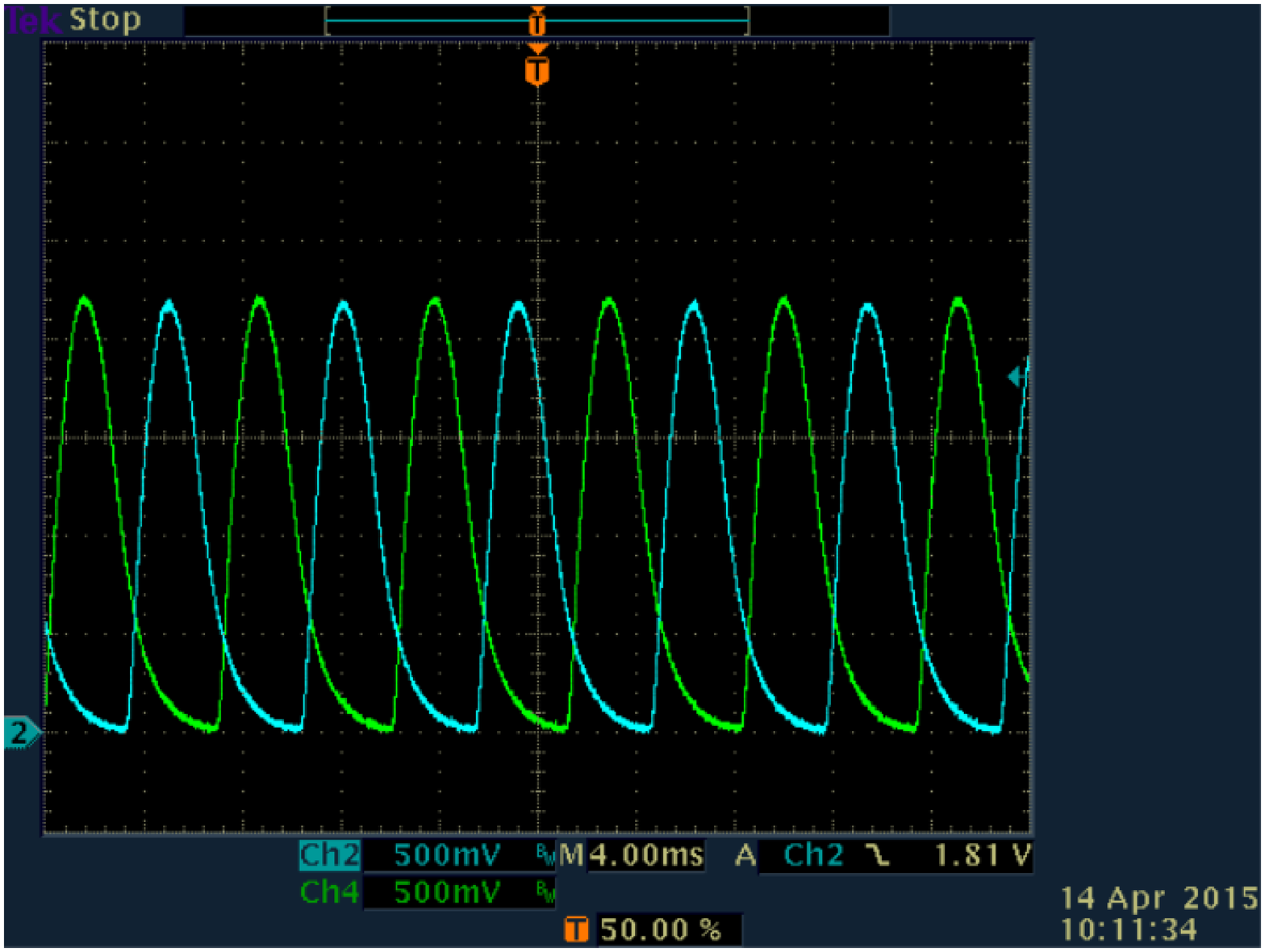}
\includegraphics[width=0.4\textwidth]{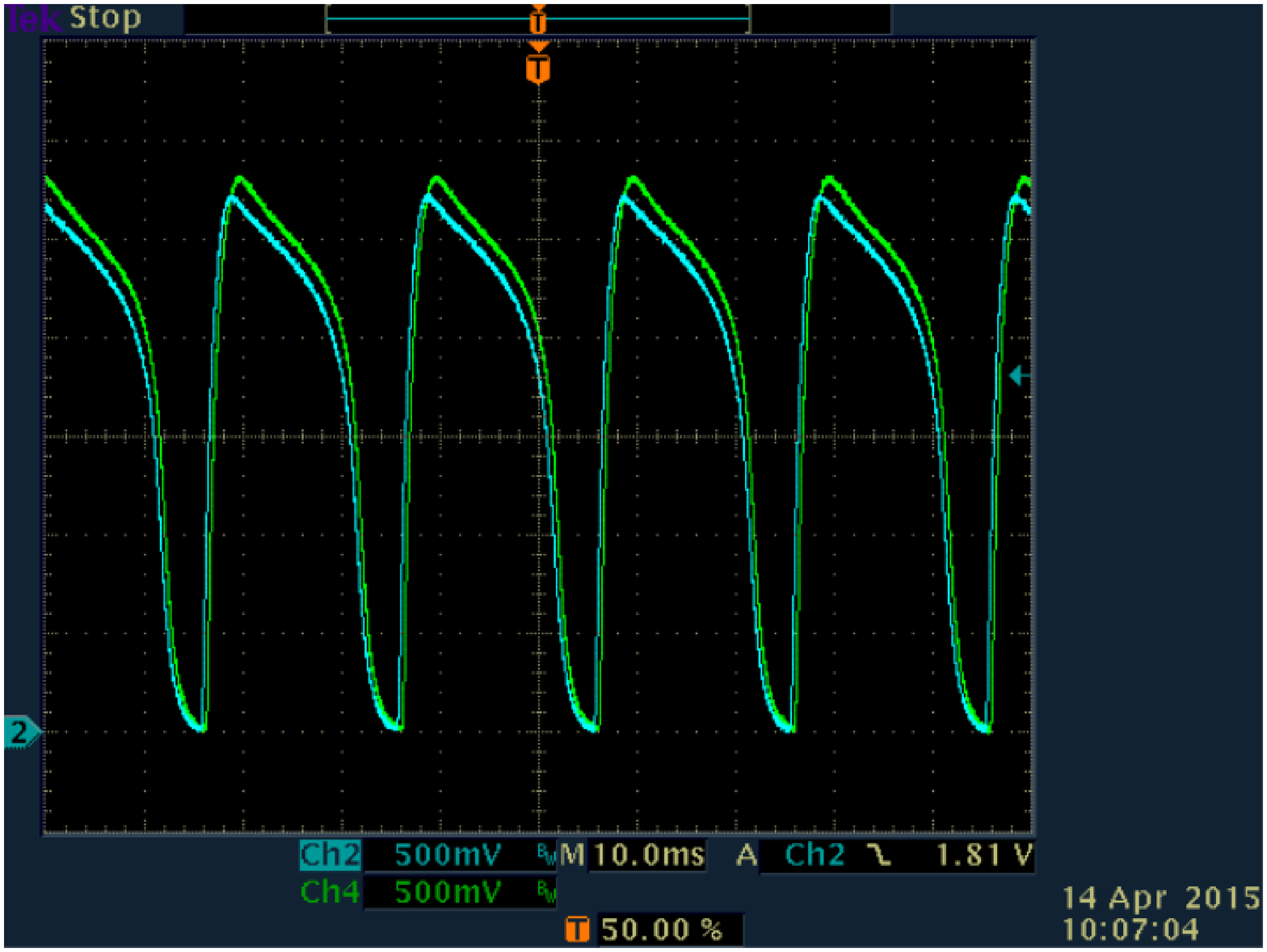}
\caption{\label{fig:AP}Screenshots of anti-phase (left) and in-phase (right) oscillations for two QS-coupled Repressilator circuits. The protein $B$ voltages from each Repressilator circuit are shown. Both screenshots use the same circuit values, thus demonstrating the coexistence of AP and IP states.}
\end{figure}

The motivation for the circuit improvements presented here is to extend our previous investigations to coupled Repressilators.  Figure \ref{fig:AP} shows examples of the oscillations for two coupled Repressilators; more extensive investigation results are in preparation. The screen-shots show the $B$-protein voltages of the two Repressilator circuits. The coupling scheme in Eqs.\ \eqref{rep-eqns} produces a multistable system whose stable oscillations are predominantly anti-phase (AP)\cite{ullner2008} like those in the top screen-shot of Fig.~\ref{fig:AP}.  Interestingly, under appropriate parameter values it is possible to find stable in-phase oscillations (IP) like those in the bottom screen-shot of Fig.~\ref{fig:AP}, which coexist with AP. Both screen-shots use the same parameters ($n=4$, $\alpha=143$, $\beta_i=0.5,0.1,0.1$, $\kappa=4.8$, $R_{hill}=9$ k$\Omega$) and both the AP and IP can be accessed simply by smoothly varying the coupling strength $Q$. AP is the sole stable state at small $Q$ and as $Q$ is increased the amplitude of the AP decreases until the AP becomes unstable and transitions to a stable steady-state characterized by both $B$-proteins being at the high value. When $Q$ is then decreased there is a transition to IP at the endpoint of the stable steady-state (similar to the decreasing Q induced transition for the single Repressilator in Fig.~\ref{fig:SS_to_LC}).   

\section{Incorporation of Additive Noise}
\begin{figure}
\includegraphics[width=0.5\textwidth]{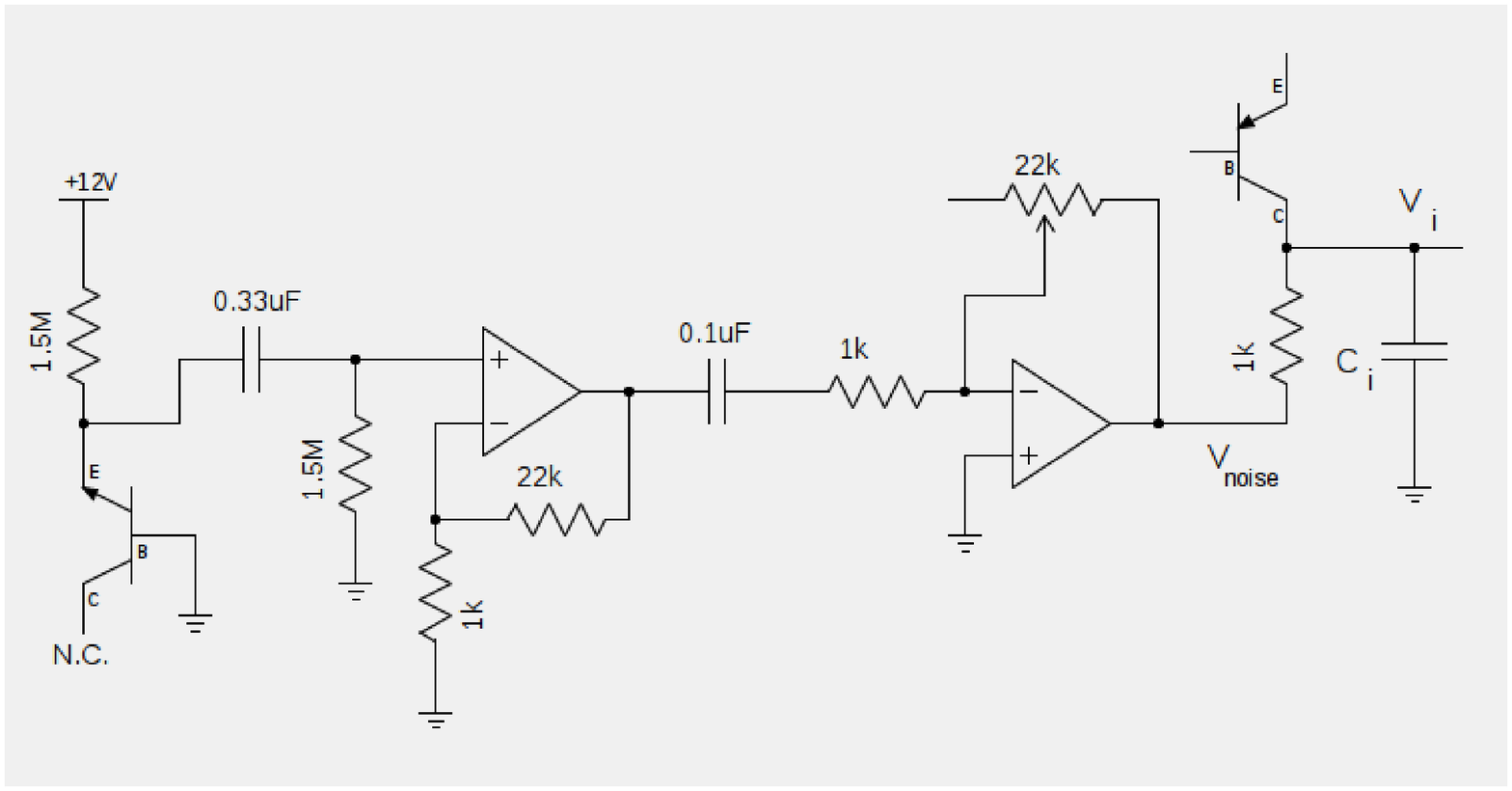}
\caption{\label{fig:noise_circuit}Noise circuit. 2N3904 \textit{npn} on left has no connection at its collector. 2N3906 \textit{pnp} on right is from the gene circuit. Op-amps are OPA228 powered by $\pm 12$ V. }
\end{figure}

Additive noise may be included using a simple noise circuit shown in Fig.~\ref{fig:noise_circuit} based on the breakdown of a reverse biased base-emitter junction as described previously.\cite{hellen2013b}  Noise is added to a protein by disconnecting its $R_C$ from ground in Fig.~\ref{fig:gene-circuit} and connecting it to the noise circuit output as shown in Fig.~\ref{fig:noise_circuit}. The potentiometer at the second op-amp adjusts the noise amplitude. Using the same procedure used to find Eq.\ \eqref{gene-circ-eqn}, the equation for the gene's protein voltage $V_i$ is easily found to be
\begin{equation}
(1k\Omega)C_i\frac{dV_i}{dt}=V_i-V_{noise}+I_t(1k\Omega).
\end{equation}
The noise is symmetric about zero and therefore the minus sign is irrelevant, thus accomplishing the task of adding noise to the protein voltage $V_i$. 

Comparison of the noise-influenced dynamical results from circuit measurements and numerical predictions requires careful connection of the electronically generated noise characteristics to the simulated noise. Here we summarize those connections, which were derived previously.\cite{hellen2013b} In simulations additive noise is typically represented by $D\eta(t)$ where $\eta(t)$ is a zero-mean Gaussian noise with unit variance and the amplitude $D$ is the noise strength. The electronically generated noise is characterized by its $rms$ amplitude $V_{Nrms}$ and its frequency bandwidth $f_c$. The relation between the simulated noise strength $D$ and the measured strength $V_{Nrms}$ is\cite{hellen2013b}
\begin{equation}
D=\frac{V_{Nrms}}{V_{th}\sqrt{\gamma (RC)f_c}}
\label{noise_strength}
\end{equation}
where $RC$ is the characteristic time of the Repressilator, and $\pi/4 \leq \gamma \leq \pi/2$ depending on the gain of the second amplifier in Fig.~\ref{fig:noise_circuit}. The noise bandwidth is determined by the op-amp's gain-bandwidth product (33 MHz for the OPA228) and the gain of the non-inverting amplifier in Fig.~\ref{fig:noise_circuit} (about $20\times$) resulting in a noise bandwidth of $f_c=33/20\approx 1.5$ MHz.

\begin{figure}
\includegraphics[width=0.475\textwidth]{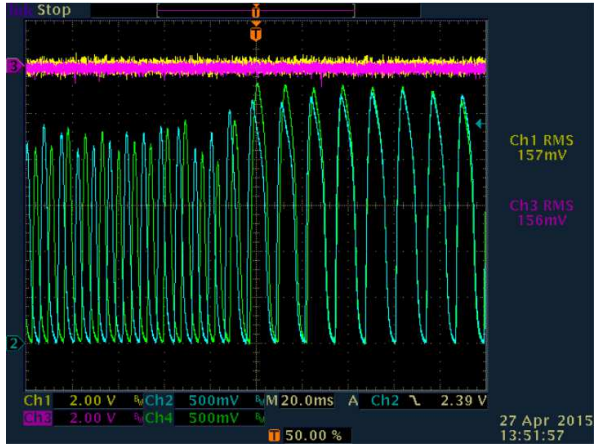}
\caption{\label{fig:nz_AP_IP}Screen-shot of noise-induced transition from anti-phase (AP) to in-phase (IP) oscillation for two QS-coupled Repressilator circuits. Independent noise (top two traces) was added to each $B$-protein voltage. }
\end{figure}

As a demonstration, we add independent noises to each $B$-protein for the case of coexistence of AP and IP states used for Fig.~\ref{fig:AP} ($n=4$, $\alpha=143$, $\kappa=4.8$).  Multiple transitions between the states were observed. Figure \ref{fig:nz_AP_IP} shows a noise-induced transition from AP to IP.  The top two traces are the added noises with $rms$-amplitudes of $V_{Nrms}=0.156$ V. Equation \eqref{noise_strength} gives the corresponding noise strength for simulation $D\approx0.6$, found using $\alpha=143$ and $I_{max}R_C=2.95$ V to give $V_{th}=20.6$ mV, characteristic time $RC=0.1$ ms, and taking $\gamma=1$.   

\section{Conclusion}
We presented a revised design for our electronic circuit model of a synthetic genetic network comprised of Repressilators coupled together by quorum sensing. Connections between mathematical parameters and circuit values were improved, in part, by including the large-signal transistor model in the derivation. The all-new quorum sensing circuitry allowed expansion of the quorum sensing circuit's accessible parameter range to match that of the mathematical model.  Important features include the incorporation of Hill function binding kinetics and the ability to set initial conditions.  Circuit behavior was verified by comparing bifurcation diagrams obtained from measurements and numerical simulation.  The circuit revisions were important because they allow us to extend previous investigations to the case of coupled Repressilators. An example of this extension demonstrated the coexistence of IP and AP oscillatory states, and noise-induced transitions between these states.  A more extensive investigation of the coupled Represilators is undertaken and to be presented in the future.  

\begin{acknowledgments}
S.K.D. acknowledges support by the CSIR Emeritus Scientist scheme. The authors thank Evgeny Volkov for valuable contributions. 
\end{acknowledgments}

\bibliography{sgn_hellen16}

\begin{thebibliography}{19}%
\makeatletter
\providecommand \@ifxundefined [1]{%
 \@ifx{#1\undefined}
}%
\providecommand \@ifnum [1]{%
 \ifnum #1\expandafter \@firstoftwo
 \else \expandafter \@secondoftwo
 \fi
}%
\providecommand \@ifx [1]{%
 \ifx #1\expandafter \@firstoftwo
 \else \expandafter \@secondoftwo
 \fi
}%
\providecommand \natexlab [1]{#1}%
\providecommand \enquote  [1]{``#1''}%
\providecommand \bibnamefont  [1]{#1}%
\providecommand \bibfnamefont [1]{#1}%
\providecommand \citenamefont [1]{#1}%
\providecommand \href@noop [0]{\@secondoftwo}%
\providecommand \href [0]{\begingroup \@sanitize@url \@href}%
\providecommand \@href[1]{\@@startlink{#1}\@@href}%
\providecommand \@@href[1]{\endgroup#1\@@endlink}%
\providecommand \@sanitize@url [0]{\catcode `\\12\catcode `\$12\catcode
  `\&12\catcode `\#12\catcode `\^12\catcode `\_12\catcode `\%12\relax}%
\providecommand \@@startlink[1]{}%
\providecommand \@@endlink[0]{}%
\providecommand \url  [0]{\begingroup\@sanitize@url \@url }%
\providecommand \@url [1]{\endgroup\@href {#1}{\urlprefix }}%
\providecommand \urlprefix  [0]{URL }%
\providecommand \Eprint [0]{\href }%
\providecommand \doibase [0]{http://dx.doi.org/}%
\providecommand \selectlanguage [0]{\@gobble}%
\providecommand \bibinfo  [0]{\@secondoftwo}%
\providecommand \bibfield  [0]{\@secondoftwo}%
\providecommand \translation [1]{[#1]}%
\providecommand \BibitemOpen [0]{}%
\providecommand \bibitemStop [0]{}%
\providecommand \bibitemNoStop [0]{.\EOS\space}%
\providecommand \EOS [0]{\spacefactor3000\relax}%
\providecommand \BibitemShut  [1]{\csname bibitem#1\endcsname}%
\let\auto@bib@innerbib\@empty
\bibitem [{\citenamefont {Hasty}\ \emph {et~al.}(2001)\citenamefont {Hasty},
  \citenamefont {Isaacs}, \citenamefont {Dolnik}, \citenamefont {McMillen},\
  and\ \citenamefont {Collins}}]{hasty2001}%
  \BibitemOpen
  \bibfield  {author} {\bibinfo {author} {\bibfnamefont {J.}~\bibnamefont
  {Hasty}}, \bibinfo {author} {\bibfnamefont {F.}~\bibnamefont {Isaacs}},
  \bibinfo {author} {\bibfnamefont {M.}~\bibnamefont {Dolnik}}, \bibinfo
  {author} {\bibfnamefont {D.}~\bibnamefont {McMillen}}, \ and\ \bibinfo
  {author} {\bibfnamefont {J.~J.}\ \bibnamefont {Collins}},\ }\bibfield
  {title} {\enquote {\bibinfo {title} {Designer gene networks: Towards
  fundamental cellular control},}\ }\href {\doibase 10.1063/1.1345702}
  {\bibfield  {journal} {\bibinfo  {journal} {Chaos: An Interdisciplinary
  Journal of Nonlinear Science}\ }\textbf {\bibinfo {volume} {11}},\ \bibinfo
  {pages} {207--220} (\bibinfo {year} {2001})}\BibitemShut {NoStop}%
\bibitem [{\citenamefont {Elowitz}\ and\ \citenamefont
  {Lim}(2010)}]{elowitz2010}%
  \BibitemOpen
  \bibfield  {author} {\bibinfo {author} {\bibfnamefont {M.}~\bibnamefont
  {Elowitz}}\ and\ \bibinfo {author} {\bibfnamefont {W.~A.}\ \bibnamefont
  {Lim}},\ }\bibfield  {title} {\enquote {\bibinfo {title} {Build life to
  understand it},}\ }\href {\doibase 10.1038/468889a} {\bibfield  {journal}
  {\bibinfo  {journal} {Nature}\ }\textbf {\bibinfo {volume} {468}},\ \bibinfo
  {pages} {889--890} (\bibinfo {year} {2010})}\BibitemShut {NoStop}%
\bibitem [{\citenamefont {Benenson}(2012)}]{benenson2012}%
  \BibitemOpen
  \bibfield  {author} {\bibinfo {author} {\bibfnamefont {Y.}~\bibnamefont
  {Benenson}},\ }\bibfield  {title} {\enquote {\bibinfo {title} {Biomolecular
  computing systems: principles, progress and potential},}\ }\href {\doibase
  10.1038/nrg3197} {\bibfield  {journal} {\bibinfo  {journal} {Nat Rev Genet}\
  }\textbf {\bibinfo {volume} {13}},\ \bibinfo {pages} {455--468} (\bibinfo
  {year} {2012})}\BibitemShut {NoStop}%
\bibitem [{\citenamefont {Gardner}, \citenamefont {Cantor},\ and\ \citenamefont
  {Collins}(2000)}]{gardner2000}%
  \BibitemOpen
  \bibfield  {author} {\bibinfo {author} {\bibfnamefont {T.~S.}\ \bibnamefont
  {Gardner}}, \bibinfo {author} {\bibfnamefont {C.~R.}\ \bibnamefont {Cantor}},
  \ and\ \bibinfo {author} {\bibfnamefont {J.~J.}\ \bibnamefont {Collins}},\
  }\bibfield  {title} {\enquote {\bibinfo {title} {Construction of a genetic
  toggle switch in escherichia coli},}\ }\href {\doibase 10.1038/35002131}
  {\bibfield  {journal} {\bibinfo  {journal} {Nature}\ }\textbf {\bibinfo
  {volume} {403}},\ \bibinfo {pages} {339--342} (\bibinfo {year}
  {2000})}\BibitemShut {NoStop}%
\bibitem [{\citenamefont {Elowitz}\ and\ \citenamefont
  {Leibler}(2000)}]{elowitz2000}%
  \BibitemOpen
  \bibfield  {author} {\bibinfo {author} {\bibfnamefont {M.~B.}\ \bibnamefont
  {Elowitz}}\ and\ \bibinfo {author} {\bibfnamefont {S.}~\bibnamefont
  {Leibler}},\ }\bibfield  {title} {\enquote {\bibinfo {title} {A synthetic
  oscillatory network of transcriptional regulators},}\ }\href {\doibase
  10.1038/35002125} {\bibfield  {journal} {\bibinfo  {journal} {Nature}\
  }\textbf {\bibinfo {volume} {403}},\ \bibinfo {pages} {335--338} (\bibinfo
  {year} {2000})}\BibitemShut {NoStop}%
\bibitem [{\citenamefont {Mason}\ \emph {et~al.}(2004)\citenamefont {Mason},
  \citenamefont {Linsay}, \citenamefont {Collins},\ and\ \citenamefont
  {Glass}}]{mason2004}%
  \BibitemOpen
  \bibfield  {author} {\bibinfo {author} {\bibfnamefont {J.}~\bibnamefont
  {Mason}}, \bibinfo {author} {\bibfnamefont {P.~S.}\ \bibnamefont {Linsay}},
  \bibinfo {author} {\bibfnamefont {J.~J.}\ \bibnamefont {Collins}}, \ and\
  \bibinfo {author} {\bibfnamefont {L.}~\bibnamefont {Glass}},\ }\bibfield
  {title} {\enquote {\bibinfo {title} {Evolving complex dynamics in electronic
  models of genetic networks},}\ }\href {\doibase
  http://dx.doi.org/10.1063/1.1786683} {\bibfield  {journal} {\bibinfo
  {journal} {Chaos: An Interdisciplinary Journal of Nonlinear Science}\
  }\textbf {\bibinfo {volume} {14}},\ \bibinfo {pages} {707--715} (\bibinfo
  {year} {2004})}\BibitemShut {NoStop}%
\bibitem [{\citenamefont {Wagemakers}\ \emph {et~al.}(2006)\citenamefont
  {Wagemakers}, \citenamefont {Buld\'{u}}, \citenamefont {Garc\'{i}a-Ojalvo},\
  and\ \citenamefont {Sanju\'{a}n}}]{wagemakers2006}%
  \BibitemOpen
  \bibfield  {author} {\bibinfo {author} {\bibfnamefont {A.}~\bibnamefont
  {Wagemakers}}, \bibinfo {author} {\bibfnamefont {J.~M.}\ \bibnamefont
  {Buld\'{u}}}, \bibinfo {author} {\bibfnamefont {J.}~\bibnamefont
  {Garc\'{i}a-Ojalvo}}, \ and\ \bibinfo {author} {\bibfnamefont {M.~A.~F.}\
  \bibnamefont {Sanju\'{a}n}},\ }\bibfield  {title} {\enquote {\bibinfo {title}
  {Synchronization of electronic genetic networks},}\ }\href {\doibase
  http://dx.doi.org/10.1063/1.2173048} {\bibfield  {journal} {\bibinfo
  {journal} {Chaos: An Interdisciplinary Journal of Nonlinear Science}\
  }\textbf {\bibinfo {volume} {16}},\ \bibinfo {eid} {013127} (\bibinfo {year}
  {2006})}\BibitemShut {NoStop}%
\bibitem [{\citenamefont {Buld\'{u}}\ \emph {et~al.}(2007)\citenamefont
  {Buld\'{u}}, \citenamefont {Garc\'{i}a-Ojalvo}, \citenamefont {Wagemakers},\
  and\ \citenamefont {Sanju\'{a}n}}]{buldu2007}%
  \BibitemOpen
  \bibfield  {author} {\bibinfo {author} {\bibfnamefont {J.~M.}\ \bibnamefont
  {Buld\'{u}}}, \bibinfo {author} {\bibfnamefont {J.}~\bibnamefont
  {Garc\'{i}a-Ojalvo}}, \bibinfo {author} {\bibfnamefont {A.}~\bibnamefont
  {Wagemakers}}, \ and\ \bibinfo {author} {\bibfnamefont {M.~A.~F.}\
  \bibnamefont {Sanju\'{a}n}},\ }\bibfield  {title} {\enquote {\bibinfo {title}
  {Electronic design of synthetic genetic networks},}\ }\href {\doibase
  10.1142/S0218127407019275} {\bibfield  {journal} {\bibinfo  {journal}
  {International Journal of Bifurcation and Chaos}\ }\textbf {\bibinfo {volume}
  {17}},\ \bibinfo {pages} {3507--3511} (\bibinfo {year} {2007})},\ \Eprint
  {http://arxiv.org/abs/http://www.worldscientific.com/doi/pdf/10.1142/S0218127407019275}
  {http://www.worldscientific.com/doi/pdf/10.1142/S0218127407019275}
  \BibitemShut {NoStop}%
\bibitem [{\citenamefont {Tokuda}, \citenamefont {Wagemakers},\ and\
  \citenamefont {Sanju\'{a}n}(2010)}]{tokuda2010}%
  \BibitemOpen
  \bibfield  {author} {\bibinfo {author} {\bibfnamefont {I.~T.}\ \bibnamefont
  {Tokuda}}, \bibinfo {author} {\bibfnamefont {A.}~\bibnamefont {Wagemakers}},
  \ and\ \bibinfo {author} {\bibfnamefont {M.~A.~F.}\ \bibnamefont
  {Sanju\'{a}n}},\ }\bibfield  {title} {\enquote {\bibinfo {title} {Predicting
  the synchronization of a network of electronic repressilators},}\ }\href
  {\doibase 10.1142/S0218127410026800} {\bibfield  {journal} {\bibinfo
  {journal} {International Journal of Bifurcation and Chaos}\ }\textbf
  {\bibinfo {volume} {20}},\ \bibinfo {pages} {1751--1760} (\bibinfo {year}
  {2010})},\ \Eprint
  {http://arxiv.org/abs/http://www.worldscientific.com/doi/pdf/10.1142/S0218127410026800}
  {http://www.worldscientific.com/doi/pdf/10.1142/S0218127410026800}
  \BibitemShut {NoStop}%
\bibitem [{\citenamefont {Hellen}\ \emph {et~al.}(2011)\citenamefont {Hellen},
  \citenamefont {Volkov}, \citenamefont {Kurths},\ and\ \citenamefont
  {Dana}}]{hellen2011}%
  \BibitemOpen
  \bibfield  {author} {\bibinfo {author} {\bibfnamefont {E.~H.}\ \bibnamefont
  {Hellen}}, \bibinfo {author} {\bibfnamefont {E.}~\bibnamefont {Volkov}},
  \bibinfo {author} {\bibfnamefont {J.}~\bibnamefont {Kurths}}, \ and\ \bibinfo
  {author} {\bibfnamefont {S.~K.}\ \bibnamefont {Dana}},\ }\bibfield  {title}
  {\enquote {\bibinfo {title} {An electronic analog of synthetic genetic
  networks},}\ }\href {\doibase 10.1371/journal.pone.0023286} {\bibfield
  {journal} {\bibinfo  {journal} {PLoS ONE}\ }\textbf {\bibinfo {volume} {6}},\
  \bibinfo {pages} {e23286} (\bibinfo {year} {2011})}\BibitemShut {NoStop}%
\bibitem [{\citenamefont {Hellen}\ \emph
  {et~al.}(2013{\natexlab{a}})\citenamefont {Hellen}, \citenamefont {Dana},
  \citenamefont {Zhurov},\ and\ \citenamefont {Volkov}}]{hellen2013}%
  \BibitemOpen
  \bibfield  {author} {\bibinfo {author} {\bibfnamefont {E.~H.}\ \bibnamefont
  {Hellen}}, \bibinfo {author} {\bibfnamefont {S.~K.}\ \bibnamefont {Dana}},
  \bibinfo {author} {\bibfnamefont {B.}~\bibnamefont {Zhurov}}, \ and\ \bibinfo
  {author} {\bibfnamefont {E.}~\bibnamefont {Volkov}},\ }\bibfield  {title}
  {\enquote {\bibinfo {title} {Electronic implementation of a repressilator
  with quorum sensing feedback},}\ }\href {\doibase
  10.1371/journal.pone.0062997} {\bibfield  {journal} {\bibinfo  {journal}
  {PLoS ONE}\ }\textbf {\bibinfo {volume} {8}},\ \bibinfo {pages} {e62997}
  (\bibinfo {year} {2013}{\natexlab{a}})}\BibitemShut {NoStop}%
\bibitem [{\citenamefont {Hellen}\ \emph
  {et~al.}(2013{\natexlab{b}})\citenamefont {Hellen}, \citenamefont {Dana},
  \citenamefont {Kurths}, \citenamefont {Kehler},\ and\ \citenamefont
  {Sinha}}]{hellen2013b}%
  \BibitemOpen
  \bibfield  {author} {\bibinfo {author} {\bibfnamefont {E.~H.}\ \bibnamefont
  {Hellen}}, \bibinfo {author} {\bibfnamefont {S.~K.}\ \bibnamefont {Dana}},
  \bibinfo {author} {\bibfnamefont {J.}~\bibnamefont {Kurths}}, \bibinfo
  {author} {\bibfnamefont {E.}~\bibnamefont {Kehler}}, \ and\ \bibinfo {author}
  {\bibfnamefont {S.}~\bibnamefont {Sinha}},\ }\bibfield  {title} {\enquote
  {\bibinfo {title} {Noise-aided logic in an electronic analog of synthetic
  genetic networks},}\ }\href {\doibase 10.1371/journal.pone.0076032}
  {\bibfield  {journal} {\bibinfo  {journal} {PLoS ONE}\ }\textbf {\bibinfo
  {volume} {8}},\ \bibinfo {pages} {e76032} (\bibinfo {year}
  {2013}{\natexlab{b}})}\BibitemShut {NoStop}%
\bibitem [{\citenamefont {Garc\'{i}a-Ojalvo}, \citenamefont {Elowitz},\ and\
  \citenamefont {Strogatz}(2004)}]{Garcia-Ojalvo2004}%
  \BibitemOpen
  \bibfield  {author} {\bibinfo {author} {\bibfnamefont {J.}~\bibnamefont
  {Garc\'{i}a-Ojalvo}}, \bibinfo {author} {\bibfnamefont {M.~B.}\ \bibnamefont
  {Elowitz}}, \ and\ \bibinfo {author} {\bibfnamefont {S.~H.}\ \bibnamefont
  {Strogatz}},\ }\bibfield  {title} {\enquote {\bibinfo {title} {Modeling a
  synthetic multicellular clock: Repressilators coupled by quorum sensing},}\
  }\href {\doibase 10.1073/pnas.0307095101} {\bibfield  {journal} {\bibinfo
  {journal} {Proceedings of the National Academy of Sciences of the United
  States of America}\ }\textbf {\bibinfo {volume} {101}},\ \bibinfo {pages}
  {10955--10960} (\bibinfo {year} {2004})},\ \Eprint
  {http://arxiv.org/abs/http://www.pnas.org/content/101/30/10955.full.pdf+html}
  {http://www.pnas.org/content/101/30/10955.full.pdf+html} \BibitemShut
  {NoStop}%
\bibitem [{\citenamefont {Potapov}, \citenamefont {Zhurov},\ and\ \citenamefont
  {Volkov}(2012)}]{potapov2012}%
  \BibitemOpen
  \bibfield  {author} {\bibinfo {author} {\bibfnamefont {I.}~\bibnamefont
  {Potapov}}, \bibinfo {author} {\bibfnamefont {B.}~\bibnamefont {Zhurov}}, \
  and\ \bibinfo {author} {\bibfnamefont {E.}~\bibnamefont {Volkov}},\
  }\bibfield  {title} {\enquote {\bibinfo {title} {“quorum sensing”
  generated multistability and chaos in a synthetic genetic oscillator},}\
  }\href {\doibase http://dx.doi.org/10.1063/1.4705085} {\bibfield  {journal}
  {\bibinfo  {journal} {Chaos: An Interdisciplinary Journal of Nonlinear
  Science}\ }\textbf {\bibinfo {volume} {22}},\ \bibinfo {eid} {023117}
  (\bibinfo {year} {2012})}\BibitemShut {NoStop}%
\bibitem [{\citenamefont {Ullner}\ \emph {et~al.}(2007)\citenamefont {Ullner},
  \citenamefont {Zaikin}, \citenamefont {Volkov},\ and\ \citenamefont
  {Garc\'{i}a-Ojalvo}}]{ullner2007}%
  \BibitemOpen
  \bibfield  {author} {\bibinfo {author} {\bibfnamefont {E.}~\bibnamefont
  {Ullner}}, \bibinfo {author} {\bibfnamefont {A.}~\bibnamefont {Zaikin}},
  \bibinfo {author} {\bibfnamefont {E.~I.}\ \bibnamefont {Volkov}}, \ and\
  \bibinfo {author} {\bibfnamefont {J.}~\bibnamefont {Garc\'{i}a-Ojalvo}},\
  }\bibfield  {title} {\enquote {\bibinfo {title} {Multistability and
  clustering in a population of synthetic genetic oscillators via
  phase-repulsive cell-to-cell communication},}\ }\href {\doibase
  10.1103/PhysRevLett.99.148103} {\bibfield  {journal} {\bibinfo  {journal}
  {Phys. Rev. Lett.}\ }\textbf {\bibinfo {volume} {99}},\ \bibinfo {pages}
  {148103} (\bibinfo {year} {2007})}\BibitemShut {NoStop}%
\bibitem [{\citenamefont {Ullner}\ \emph {et~al.}(2008)\citenamefont {Ullner},
  \citenamefont {Koseska}, \citenamefont {Kurths}, \citenamefont {Volkov},
  \citenamefont {Kantz},\ and\ \citenamefont {Garc\'{i}a-Ojalvo}}]{ullner2008}%
  \BibitemOpen
  \bibfield  {author} {\bibinfo {author} {\bibfnamefont {E.}~\bibnamefont
  {Ullner}}, \bibinfo {author} {\bibfnamefont {A.}~\bibnamefont {Koseska}},
  \bibinfo {author} {\bibfnamefont {J.}~\bibnamefont {Kurths}}, \bibinfo
  {author} {\bibfnamefont {E.}~\bibnamefont {Volkov}}, \bibinfo {author}
  {\bibfnamefont {H.}~\bibnamefont {Kantz}}, \ and\ \bibinfo {author}
  {\bibfnamefont {J.}~\bibnamefont {Garc\'{i}a-Ojalvo}},\ }\bibfield  {title}
  {\enquote {\bibinfo {title} {Multistability of synthetic genetic networks
  with repressive cell-to-cell communication},}\ }\href {\doibase
  10.1103/PhysRevE.78.031904} {\bibfield  {journal} {\bibinfo  {journal} {Phys.
  Rev. E}\ }\textbf {\bibinfo {volume} {78}},\ \bibinfo {pages} {031904}
  (\bibinfo {year} {2008})}\BibitemShut {NoStop}%
\bibitem [{\citenamefont {Strogatz}\ and\ \citenamefont
  {Stewart}(1993)}]{strogatz1993}%
  \BibitemOpen
  \bibfield  {author} {\bibinfo {author} {\bibfnamefont {S.~H.}\ \bibnamefont
  {Strogatz}}\ and\ \bibinfo {author} {\bibfnamefont {I.}~\bibnamefont
  {Stewart}},\ }\bibfield  {title} {\enquote {\bibinfo {title} {Coupled
  oscillators and biological synchronization},}\ }\href@noop {} {\bibfield
  {journal} {\bibinfo  {journal} {Sci Am}\ }\textbf {\bibinfo {volume} {269}},\
  \bibinfo {pages} {102--109} (\bibinfo {year} {1993})}\BibitemShut {NoStop}%
\bibitem [{\citenamefont {Potapov}, \citenamefont {Volkov},\ and\ \citenamefont
  {Kuznetsov}(2011)}]{potapov2011}%
  \BibitemOpen
  \bibfield  {author} {\bibinfo {author} {\bibfnamefont {I.}~\bibnamefont
  {Potapov}}, \bibinfo {author} {\bibfnamefont {E.}~\bibnamefont {Volkov}}, \
  and\ \bibinfo {author} {\bibfnamefont {A.}~\bibnamefont {Kuznetsov}},\
  }\bibfield  {title} {\enquote {\bibinfo {title} {Dynamics of coupled
  repressilators: The role of mrna kinetics and transcription cooperativity},}\
  }\href {\doibase 10.1103/PhysRevE.83.031901} {\bibfield  {journal} {\bibinfo
  {journal} {Phys. Rev. E}\ }\textbf {\bibinfo {volume} {83}},\ \bibinfo
  {pages} {031901} (\bibinfo {year} {2011})}\BibitemShut {NoStop}%
\bibitem [{\citenamefont {Ermentrout}(2002)}]{ermentrout}%
  \BibitemOpen
  \bibfield  {author} {\bibinfo {author} {\bibfnamefont {B.}~\bibnamefont
  {Ermentrout}},\ }\href@noop {} {\emph {\bibinfo {title} {Simulating,
  Analyzing, and Animating Dynamical Systems: A Guide to XPPAUT for Researchers
  and Students}}},\ Software, Environments and Tools (Book 14)\ (\bibinfo
  {publisher} {SIAM},\ \bibinfo {year} {2002})\BibitemShut {NoStop}%
\end{thebibliography}%

\end{document}